\documentstyle[pre,aps,epsf]{revtex}
\def\la{\langle} 

\def\la{\langle}
\def\ra{\rangle}

\newcommand{\be}{\begin{eqnarray}}
\newcommand{\ee}{\end{eqnarray}}

\newcommand{\ben}{\begin{eqnaray}}
\newcommand{\een}{\end{eqnarray}}
\renewcommand{\AA}{{\cal A}}

\newcommand{\BB}{{\cal B}}
\newcommand{\ZZ}{{\cal Z}}

\newcommand{\zb}{\bar{z}}

\newcommand{\arr}[4]{
\left(\begin{array}{cc}
#1&#2\\
#3&#4
\end{array}\right)
}

\newcommand{\corr}[1]{\Big\la{#1}\Big\rangle}

\newcommand{\eq}{\begin{equation}} \newcommand{\eqx}{\end{equation}}

\newcommand{\eqn}{\begin{eqnarray}} \newcommand{\eqnx}{\end{eqnarray}}
\newcommand{\f}[2]{\frac{#1}{#2}}

\newcommand{\MM}{{\cal M}}
\newcommand{\GG}{{\cal G}}

\newcommand{\lm}{\lambda}

\newcommand{\eps}{\epsilon}

\begin{document}
\draft

\title{Correlations of Eigenvectors for Non-Hermitian Random-Matrix Models}

\author{ {\bf Romuald A. Janik}$^{1,2}$,{\bf Wolfgang
N\"{o}renberg}$^{3}$,  {\bf Maciej A.  Nowak}$^{2,3}$,
{\bf G\'{a}bor Papp}$^{4}$ and {\bf Ismail Zahed}$^5$}
\address{$^1$ Service de Physique Th\'{e}orique, CEA Saclay, 
F-91191 Gif-Sur-Yvette, France.\\
$^2$ Marian Smoluchowski Institute  of Physics, Jagellonian University, 30-059
Krakow, Poland.\\ $^3$ GSI, Planckstr. 1, D-64291 Darmstadt, Germany\\ 
$^4$CNR Department of Physics, KSU, Kent, Ohio 44242, USA \& \\
Institute for Theoretical Physics, E\"{o}tv\"{o}s University, 
Budapest, Hungary\\
$^5$Department of Physics and Astronomy,
 SUNY, Stony Brook, New York 11794, USA.}
\date{\today} \maketitle

\begin{abstract}
We establish a general relation between the diagonal correlator
of eigenvectors and the spectral  Green's function  
for non-hermitian random-matrix models in the large-$N$ 
limit. We apply this result to a number of non-hermitian random-matrix
models and show that the outcome is in good agreement with numerical 
results.

\end{abstract}
%\pacs{}

%\centerline{\bf 1. INTRODUCTION}
\section{Introduction}

Recently, a number of  new results for 
non-hermitian random-matrix (NHRM) ensembles  were
obtained~(see e.g. \cite{SOMMERS,EFETOV,HATANONELSON,US,ZEENEW1} 
and references therein), 
reflecting on the  rapidly  growing interest in  properties of NHRM in several
areas  of physics.
In a recent paper~\cite{CHALKER}, Chalker and Mehlig have pointed
out  the existence of remarkable correlations between left and right 
eigenvectors associated with  pairs of eigenvalues lying
close in the complex plane. Such effects
do not exist for hermitian (more generally normal)
random-matrix models, since in this case the left and right
eigenvectors coincide.
Some observables related to the eigenvector properties in non-hermitian
random-matrix models have been introduced and studied numerically
in~\cite{ROTTER}. 
 Chalker and Mehlig~\cite{CHALKER} obtained analytical 
formulae (in the large-$N$ limit,
where $N$ is the size of the NHRM) for correlations between left and right
eigenvectors in the case of  Ginibre's ensemble~\cite{GINIBRE}. However an
efficient calculational scheme for the simplest (and perhaps physically
 more transparent) one-point function $O(z)$ was lacking. 
In this paper we prove   a  simple relation  stating that 
the correlator between left and right
eigenvectors corresponding to the same eigenvalue is exactly 
equal, in the large-$N$ limit, to the square of the off-diagonal 
one-point Green's function~\cite{US} for non-hermitian eigenvalues.
The latter  is readily calculable for a wide variety of NHRM ensembles. 
We illustrate our observation in a number of NHRM models and show that
it agrees with numerical calculations.

The present results can be used to study the interplay between reorganization
of the left and right eigenvectors and structural changes in the complex
spectrum. In the simplest non-hermitian model -- Ginibre's  ensemble
(or generalizations thereof~\cite{GIRKO,SOMMERS88}) -- 
the density of complex 
eigenvalues is constant, filling uniformly the circle (ellipse)
in  the complex plane. 
This  model and its variants
do not have  external parameters, which could induce 
structural changes (``phase changes'') 
in the average eigenvalue distribution  e.g. 
the changes from simply to  multiple-connected domain of eigenvalues.
In order to study these more complex phenomenae one has to consider e.g.
the model for open chaotic scattering, 
where at large couplings (strong dissipation) 
the eigenvalue distribution 
splits into two disconnected islands, reflecting on the
separation of  time 
scales~\cite{WEIDMAUX,HAAK}. 
One  island corresponds to   
short-lived resonances, while  the other to  long-lived
(almost classical) trapped states~\cite{ZEL}.
It is  known in this case, that the reorganization of the
eigenvalues
is followed by some reorganization of the eigenvectors~\cite{ROTTER}. 
In particular, 
the norm of the right states is  sensitive to the distribution of
resonances~\cite{ROTTER}.

In section~2, we introduce the notations and summarize the main
results for the eigenvector correlators  in the case of 
 Ginibre's ensemble~\cite{GINIBRE}, 
as established  recently by Chalker and Mehlig~\cite{CHALKER}. 
In section~3, we present our main result and apply it to several
NHRM models with direct comparison to numerical results. Our analysis
relies on novel techniques for NHRM models discussed by some of us~\cite{US}.
A summary of our results is given in section~4, and a number of technical
details can be found in the Appendices.

%\vskip .9cm
%\centerline{\bf 2. GINIBRE  MATRIX MODEL}
%\vskip .15cm
\section{Ginibre's Matrix Model}
Ginibre~\cite{GINIBRE}
 has introduced a Gaussian ensemble of general complex matrices
$N\times N$,
i.e. matrices distributed with the probability
\be
P(\MM)d\MM \sim \exp(-N {\rm Tr}\MM \MM^{\dagger}) d\MM
\label{measure}
\ee  
giving  non-vanishing cumulants $\la\MM_{ab} {\overline{\MM}}_{ab}\ra=1/N$.
 The eigenvalues are uniformly distributed  on a unit disc centered at the
origin of the complex plane. 
In  Appendix B, we provide a short  derivation of this result 
and others using  (matrix-valued) Blue's functions~\cite{US,ZEENEW1}. 
Since the matrices are complex (non-hermitian), 
there exists a bi-orthogonal set of
right ($R$) and left ($L$) eigenvectors, so  that
\be
\MM&=&\sum_a \lambda_a |R_a><L_a| \ , \nonumber \\ 
\MM^{\dagger}&=&\sum_b \bar{\lambda}_b |L_b> <R_b|  \ ,
\label{comp}
\ee
where $<L_a|R_b>=\delta_{ab}$.
Chalker and Mehlig~\cite{CHALKER} 
have studied the following eigenvector correlators:
\be
O(z)&=&\left< \frac{1}{N} \sum_a O_{aa} \delta(z-\lambda_a) \right> \ ,
\nonumber \\
O(z,w)&=&\left< \frac{1}{N} \sum_{a \neq b}  O_{ab}
 \delta(z-\lambda_a)\delta(w-\lambda_b) \right> \ ,
 \label{cordef}
\ee
where $O_{ab}=<L_a|L_b><R_b|R_a>$. 
The main results of their paper are  exact expressions for $O(z,w)$
and for $O(z)$ in the case of   Ginibre's ensemble. 
For $N\gg1, |z-w| \neq 0$,  where $z$, $w$ are lying within the unit circle
\be
O(z,w)=-\frac{1}{\pi^2} \frac{1-z\bar{w}}{|z-w|^4} \,.
\label{wide}
\ee
For $|z-w| \rightarrow 0$, 
\be
O(z,w)_{\rm micro}=-\frac{N^2}{\pi^2} \frac{1-|Z|^2}{|\omega|^4}
[1-(1+|\omega|^2)\exp (-|\omega|^2)] \ ,
\label{narrow}
\ee 
where $Z=(z+w)/2$ and $\omega=\sqrt{N}(z-w)$.
The diagonal
correlator reads
\be 
O(z)=\frac{N}{\pi}(1-|z|^2)\,.
\label{diag}
\ee
Whereas the calculation of the ``wide'' eigenvector correlators $O(z,w)$ 
(like (\ref{wide})) in many cases of NHRM is straightforward, using e.g. 
non-hermitian diagrammatics~\cite{US}, the calculation of 
the ``close'' eigenvector correlators $O(z-w)_{micro}$ and the 
diagonal correlators $O(z)$ is technically involved.
In practice, it seems that  explicit  calculations in the microscopic 
limit  are possible 
only in the cases when the spectrum of NHRM possesses an azimuthal
(rotational) symmetry. 
In the next section we provide a general   formula
for the diagonal eigenvector correlator $O(z)$
 in terms of  the spectral one-point  Green's function.

\begin{figure}
\centerline{\epsfysize=65mm \epsfbox{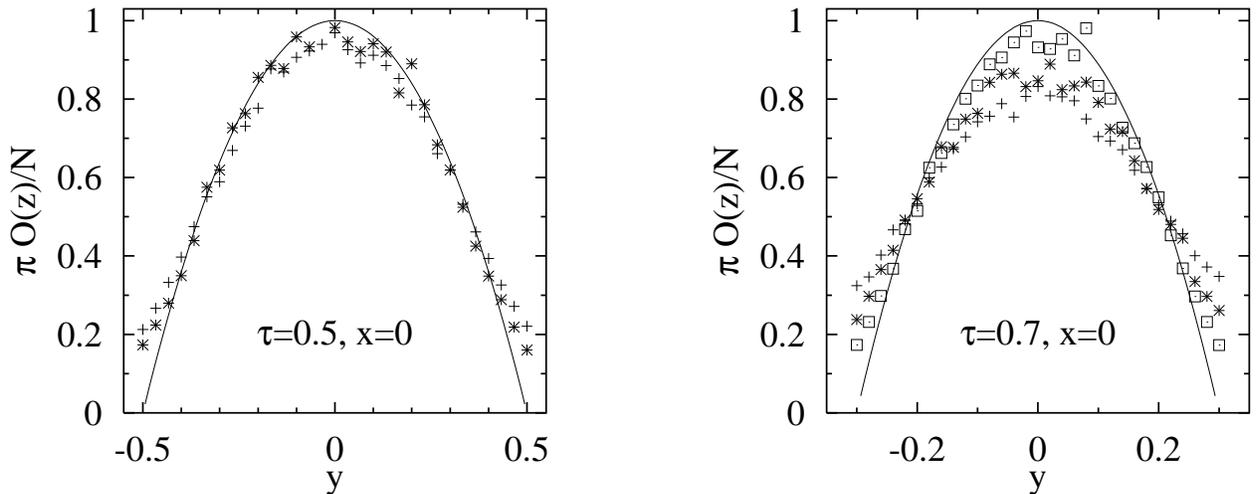}}
\caption{Eigenvector correlator $O(z=0+iy)$, generated
from an elliptic ensemble of 
  5$\cdot$10$^5$ matrices of size 20 (crosses), an ensemble of 10$^5$ matrices
of size 40 (stars) and size 100 (boxes) versus the analytical correlator
$-\GG_{\bar{q}q} \GG_{q\bar{q}}$, for $\tau=0.5$ (left) and $\tau=0.7$ (right).}
\label{girkogi}
\end{figure}

%\vskip .9cm
%\centerline{\bf 3.NEW  FORMULA and RESULTS}
%\vskip .15cm
\section{New formula and results}

The main result of this paper  reads
\be
O(z)=- \frac{N}{\pi}\GG_{q \bar{q}}\cdot \GG_{\bar{q} q} \,.
\label{hypothesis}
\ee
Here  $\GG_{q \bar{q}}$ and $\GG_{\bar{q} q} $ are off-diagonal 
elements of the generalized (2 by 2) spectral 
Green's function ${\cal G}$ ~\cite{US}
\be
{{\cal {G}}}=\arr{{\cal G}_{qq}}{{\cal G}_{q\overline{q}}}{{\cal G}_{\overline{q}q}}
{{\cal G}_{\overline{q}\overline{q}}} \,.
\label{18}
\ee
The elements ${\cal G}_{ab}$ are defined as traces ${\cal G}_{ab}=
\frac{1}{N} {\rm Tr}_N \hat{{\cal G}}_{ab}$ of the $N\times N$ blocks
of the generalized resolvent~\cite{US},
\be
\hat{{\cal {G}}}=\arr{\hat{{\cal G}}_{qq}}{\hat{{\cal
G}}_{q\overline{q}}}{
\hat{{\cal G}}_{\overline{q}q}}
{\hat{{\cal G}}_{\overline{q}\overline{q}}}
= \left\langle \arr{z-\MM}{i\epsilon {\bf 1}_N}{i\epsilon {\bf 1}_N}{\zb 
-\MM^{\dagger}}^{-1}\right\rangle \ ,
\label{19}
\ee
where ${\bf 1}_N$ is the  $N$-dimensional identity  matrix and 
 $<...>$ denotes the averaging over  
the pertinent ensemble of random matrices ${\cal M}$.
In comparison to the original work~\cite{US}, 
we have chosen here  purely imaginary infinitesimal values  in the 
off-diagonal block. This way guarantees 
that the  mathematical operations
performed in the proof of (\ref{hypothesis}) (cf. Appendix A) 
are well-defined.  

%Here $\frac{1}{N} {\rm Tr}_N \hat{{\cal G}_{ab}}={\cal G}_{ab}$, 
%for $a,b=q, \bar{q}$.
The spectral density follows 
from Gauss law~\cite{SOMMERS88},
\eq
\nu (z, \zb) = \frac 1{\pi} \partial_{\zb} \,\, {\cal G}_{qq}(z,\zb ) 
\label{20}
\eqx
which is the distribution of eigenvalues of $\MM$. 
For hermitian $\MM$, (\ref{20}) can be nonzero only on the real axis. As 
$\eps\rightarrow 0$, the block-structure decouples, and we are left
with the original resolvent. For $z\rightarrow +i0$, the latter is just a 
measurement of the real eigenvalue distribution.

 For non-hermitian $\MM$,
as $\eps\rightarrow 0$, the block
structure does not decouple, leading to a non-holomorphic resolvent for certain
two-dimensional domains on the $z$-plane.
For more technical details we refer to the original
work~\cite{US}, 
or recent reviews~\cite{NATO,BLUESHADES}.
Similar constructions have 
 been proposed recently 
in~\cite{ZEENEW1,CHALKERWANG}.

The r.h.s. of the relation~(\ref{hypothesis}) is usually given by a simple
analytical formula. 
Technically, the most efficient way of calculating 
the off-diagonal components of the Green's functions 
is to use the generalized Blue's function technique~\cite{US,ZEENEW1}.  
In  Appendix B we provide a pedagogical derivation of some of the
results below, for others we refer to the original papers. 

For  Ginibre's ensemble  
we immediately get (cf. Appendix B)
\be
O(z)_{Ginibre}= -\frac{N}{\pi}|\sqrt{z\bar{z}-1}|^2=\frac{N}{\pi}(1-|z|^2)
\ee
in agreement with Chalker and Mehlig~\cite{CHALKER}.
For the elliptic ensemble~\cite{GIRKO,SOMMERS88}, 
using the off-diagonal elements from   
Appendix B, the diagonal correlator  reads
\be
O(z)_{Elliptic}= \frac{N}{\pi} \frac1{(1-\tau^2)^2}
	\left[(1-\tau^2)^2-(1+\tau^2)|z|^2+2\tau {\rm Re} z^2\right] \,.
\label{corgir}
\ee
In Fig.~\ref{girkogi}, we present  numerical results generated from 
computer simulation of eigenvectors for the elliptic  ensemble  
with $\tau=0.5$ and $\tau=0.7$ and different size of the matrix, $N$,
versus the analytical prediction~(\ref{corgir}). The results are satisfactory.
The figure provides also a rough
 estimation of finite-size effects.
We note that numerical simulations of eigenvector correlations
are time consuming, hence the utility  of the  result
(\ref{hypothesis}), where the r.h.s is straightforward to calculate.

The models considered above have constant density of complex eigenvalues,
and the domain of eigenvalues is  simply connected (circle or ellipse). 
Below, we  consider a toy-model, where the domain 
could split into two disconnected domains at some critical value of the
external parameter. Also the distribution of  eigenvalues
is non-uniform.  The simplest non-hermitian 
model of this kind
is  Ginibre's random Hamiltonian plus a
two level deterministic Hamiltonian, with $N/2$ levels $a$
and $N/2$ levels $-a$.\footnote{To our knowledge, 
this  model was first  considered 
by Feinberg and Zee, using their  hermitization 
method~\protect~{\cite{ZEENEW1}}.}
Using the addition law for the generalized Blue's function, we 
easily obtain  all the components of the matrix valued Green's function
${\cal G}$ (cf. Appendix B).

Using  the relation (\ref{hypothesis}) we predict
\be
O(z)_{G+D}= -\frac{N}{\pi}
\left[ |z|^2+a^2-\f12(1+\sqrt{1+4a^2(z+\bar{z})^2}) \right] \,.
\label{twoplusg}
\ee

\begin{figure}
\centerline{\epsfxsize=\textwidth \epsfbox{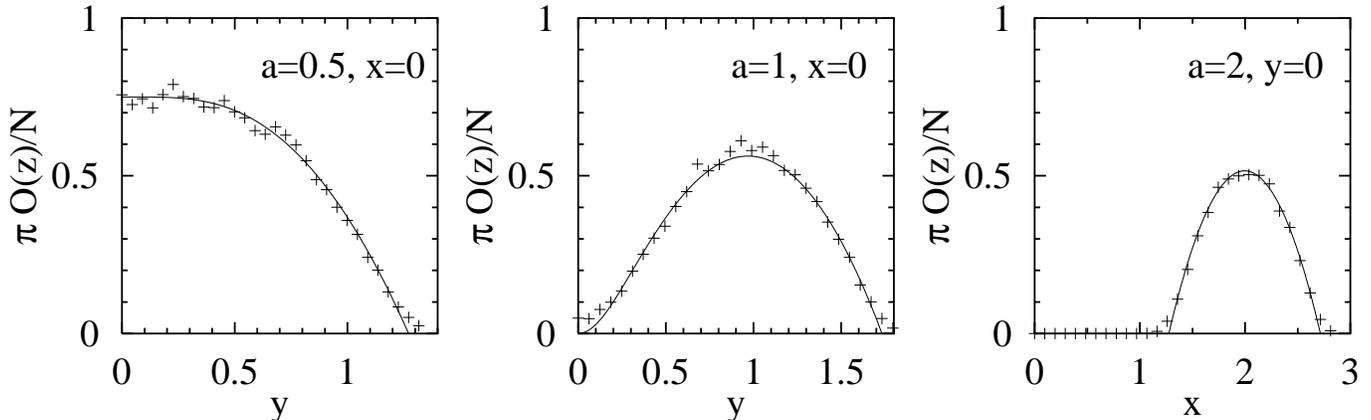}}
\caption{``Ginibre+deterministic'' eigenvector correlator $O(z=0+iy)$,
generated from an ensemble of 10$^5$ matrices of size 60 (crosses)
versus the analytical correlator $-\GG_{\bar{q}q} \GG_{q\bar{q}}$, 
at $a=0.5$ (left), $a=1$ (center), and the correlator $O(z=x+i0)$
at $a=2$ (right).}
\label{girdet}
\end{figure}

We  note  that the value $a=1$ is the critical one, 
when the single island of eigenvalues splits into two. 
Numerical simulations for this ensemble are shown in Fig.~\ref{girdet}.
The solid line is the analytical result, 
the crosses are the numerical results calculated for the line
$z=0+iy$.
The diagonal eigenvector correlators 
$O(z)$ {\it inside} the
islands follow  the distribution determined 
by the off-diagonal components for the spectral Green's function.
We also note  that the 
eigenvector correlator follows precisely   the boundary-shape 
of the eigenvalues.  
This is expected, since the 
condition $|\GG_{q\bar{q}}|=0$ determines~\cite{US} 
the regions in  the  complex plane
separating the holomorphic and non-holomorphic components of the spectral
Green's function~\footnote{The
 shape of the eigenvalue domains
for NHRM models can  be inferred from associated  hermitian models using
conformal mapping~\cite{US}. This points to yet another relationship  between
the eigenvalues and  eigenvectors of  hermitian and non-hermitian
models.}.

\begin{figure}[t]
\centerline{\epsfysize=90mm \epsfbox{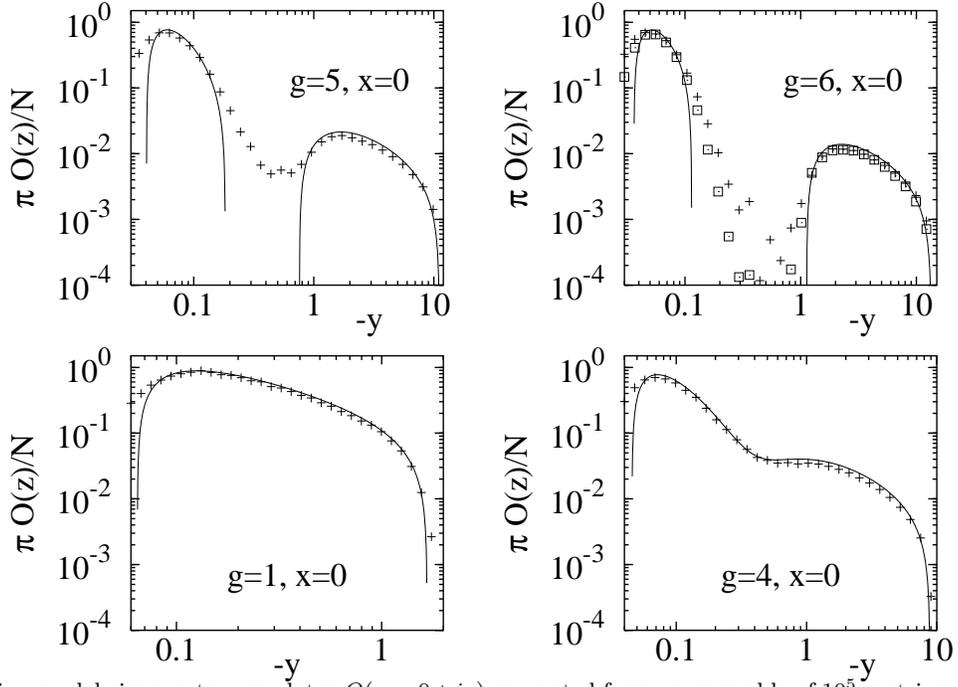}}
\caption{Scattering model eigenvector correlator $O(z=0+iy)$  generated
from an ensemble of 10$^5$ matrices of size 60 (crosses) and of size 120
(boxes) with $m=0.25$,
versus the analytical correlator $-\GG_{\bar{q}q} \GG_{q\bar{q}}$, at
different values of the coupling constant $g$.} 
\label{haake}
\end{figure}

Finally, we consider the case of open chaotic scattering. As an
illustration we choose the classical  result of Haake, Sommers and 
coworkers~\cite{HAAK}, 
based on  Mahaux-Weidenm\"{u}ller~\cite{WEIDMAUX} 
microscopic picture for nuclear reactions.
In brief, the model  is generically described by a non-hermitian
Hamiltonian of the form
\be
H-igVV^{\dagger} \ ,
\label{scat}
\ee
where $H$ is a random Gaussian (orthogonal) $N \times N$ matrix,
while $V$ is an $N \times M$ random matrix~\cite{HAAK}.
Here $N$ is a  number of  discrete states and $M<N$ is a  number of 
the continua. The model was solved~\cite{HAAK} in the limit
$N \rightarrow \infty$, $M \rightarrow \infty$, $m\equiv M/N$ fixed.   
Using the results from  Blue's function techniques (cf. Appendix B)
we predict the analytical behavior for the correlator $O(z)$ for this
model as
\be
O(z)= \f{N}{\pi} \left[ \f1{1-gy} -\f{x^2}4 - \f14
	\left( \f{g}{1-gy}+\f{m}y+\f1g \right)^2 \right]
\label{hakcor}
\ee
with $z=x+iy$.
In Fig.~\ref{haake} we compare this result to a  numerically generated
ensemble of eigenvectors. We would like to note
large finite size effects at the edges of the islands. 

Since in this model a change in the parameters $g$ and $m$ can
generate structural changes in the distribution of eigenvalues, it is
interesting to study whether the splitting of the ``islands''  is
accompanied by some distinct behavior of the eigenvector correlators.

\begin{figure}[h]
\centerline{\epsfysize=90mm \epsfbox{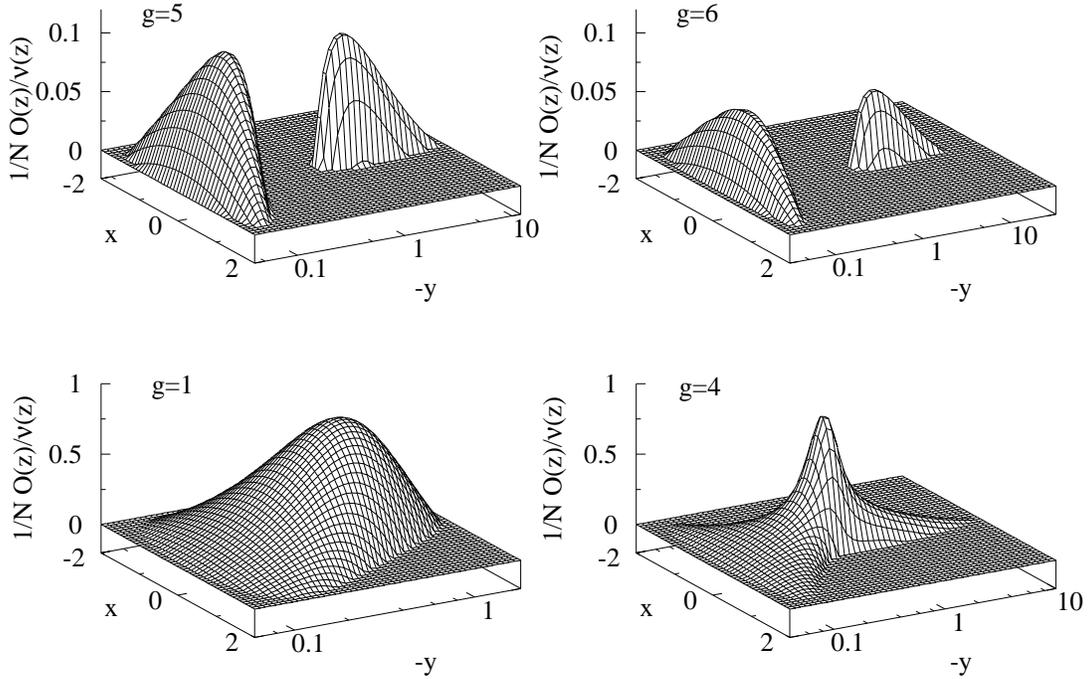}}
\caption{Three dimensional plots of the  analytical results for 
the eigenvector correlator $O(z)$, normalized by the spectral density
$\nu (z, \bar{z})$,
versus $x$  and $y$ 
for different  couplings $g$ un the case of  open chaotic scattering.}
\label{haake3}
\end{figure}

In Fig.~\ref{haake3} we plot the analytical result (\ref{hakcor}),
normalized by  the spectral density $\nu(z,\bar{z})$ (cf.  Appendix B), 
as a function of $x$ and $y$, where $z=x+iy$.
For the case considered here ($m=0.25$) the value of the critical coupling
is $g=4.44$.\footnote{Generally, 
$g^2_{\rm crit}=(1-\sqrt[3]{m})^{-3}$~\protect{\cite{HAAK}}.}
We observe a strong reorganization in the distribution of the 
average ``norm'' of 
eigenvectors
in the vicinity of the critical coupling. 

We have not taken up  in this  paper  the issue of  wide eigenvector
correlators  $O(z,w)$. For  the cases considered  here, these correlators 
are readily  constructed using the NHRM diagrammatic Bethe-Salpeter equations 
\cite{US}, as noted by Chalker and Mehlig~\cite{CHALKER}. 
In most cases, however, 
the resulting final formulae are rather lengthy.
We note that the knowledge of $O(z,w)$ in the regime $z-w\sim O(1)$ does not
suffice to determine $O(z)$ through  the sum rules~\cite{CHALKER} 
originating from the bi-orthogonality of the left/right eigenvectors.
Hence the relevance  of the present investigation.

%, despite 
%the diagonal and off-diagonal eigenvector correlators are 
%non-trivially (indirectly)  related by the sum rules~\cite{CHALKER} 
%originating from  
%the bi-orthogonality of the left/right eigenvectors.

%\vskip .9cm
%\centerline{\bf 4. SUMMARY}
%\vskip .15cm
\section{Summary}

We have established a novel relation between the diagonal 
correlator of eigenvectors and the off-diagonal elements 
of one-point spectral Green's function
for general ensembles of NHRM models.
We have appplied this result to a number of NHRM models and checked
its validity against numerically generated results. 
 Our observation accounts for part of the eigenvector
correlations established recently by a number of authors~\cite{CHALKER,ROTTER}.
 Our result generalizes to non-hermitian ensembles
with real, complex or quaternionic components, as well as non-hermitian
ensembles with additional symmetry (e.g. chiral NHRM). In this last
case, however, non-hermitian diagrammatic techniques have to be used
instead of the versatile method of Blue's functions.

Finally, we  point at the possibility of relating the eigenvector 
correlators to the eigenvalue 
correlators in the microscopic limit. This issue  and others will be discussed 
elsewhere. 

%\vskip 1.2cm
%{\bf Acknowledgments}
%\vskip 0.1cm
\acknowledgments

We  would like to thank Evgueni Kolomeitsev 
for discussions.
This work was supported in part  by the US DOE grants DE-FG-88ER40388
and DE-FG02-86ER40251,
by the Polish Government Project (KBN)  grant 
2P03B00814 and by the Hungarian grant OTKA F026622.

\appendix
%\vskip 1.5cm
%{\bf Appendix A}
%\vskip .2cm
\section{Proof of Eq. (7)}

We prove first the following representation for $O(z)$
\be 
\lim_{\eps \rightarrow 0}
\left<{\rm Tr}
\frac{\eps}{(z-\MM)(\bar{z}-\MM^{\dagger})+\eps^2}
\cdot{\rm Tr}
\frac{\eps}{(\bar{z}-\MM^{\dagger})(z-\MM)+\eps^2}\right>= \pi N\, O(z)
	\,.
\label{main}
\ee
Below we use  Tr $A=\log \det A$, and the fact that $\MM$
can be diagonalized by a {\em non-unitary} transformation $U$, 
$U^{-1}\MM U= {\rm diag}\,\,(\lambda_1,\ldots,\lambda_N)$.
Introducing $V=U^{-1}U^{{-1}\dagger}$ we get
\be
\label{e.trdet}
{\rm Tr} \frac{\eps}{(z-\MM)(\bar{z}-\MM^{\dagger})+\eps^2}=
 \frac{1}{2}\frac{d}{d \eps}\log \det [(z-\lambda_i)V_{ik}(\bar{z}-
\bar{\lambda}_k) +\eps^2 V_{ik}] \equiv \frac{1}{2}\partial_{\eps} \log
\det [\cdots] \,.
\ee
Using this representation, we note that $O(z)$ is zero in the limiting
procedure,
 unless 
$z$ is close to any (say $\lm_1$) of the eigenvalues of $\MM$. If this
happens, we use the parameterization, 
$z-\lambda_1=\eps u \exp (i\phi)$ 
with $u \sim O(1)$. Of course we get additional (similar)
contributions when $z$ is close to the other eigenvalues $\lm_i$. For
notational 
simplicity we will now consider only the case of $\lm_1$, adding the
remaining contributions in (\ref{crucial}). 

Using  Laplace's expansion for the first row of  the determinant in
(\ref{e.trdet}) we arrive in the ${\eps \rightarrow 0}$ limit
\be
\frac{1}{2}\partial_{\eps} \log \det [\cdots] =
\frac{1}{\eps} \frac{1}{1+u^2 \frac{\det W_2}{V_{11} \det
W_1}}
	\ ,
\ee
with $W_1$ being the matrix
$(z-\lambda_i)V_{ik}(\bar{z}-\bar{\lambda}_k)$ without the first row and
column and $(W_2)_{i+1,k+1}=(W_1)_{ik}$,
$(W_2)_{1,k}=(\bar{z}-\bar{\lambda}_k) V_{1k}$, 
$(W_2)_{i,1}=(z-\lambda_k) V_{i1}$ and $(W_2)_{11}=V_{11}$.

{}From Laplace's expansion
\be
\det W_1&=& \det V' \prod_{i=2,N}
(z-\lambda_i)(\bar{z}-\bar{\lambda}_i)\ , \nonumber\\
\det W_2&=& \det V \prod_{i=2,N} (z-\lambda_i)(\bar{z}-\bar{\lambda}_i)
\ee
the ratio of the determinants $\det W_1/\det W_2$ is simply $(V^{-1})_{11}$.
In the above, $V'$ is the minor of  $V$ left after crossing out the first
row and first column. In this way we obtain:
\eq
\frac{1}{2}\partial_{\eps} \log \det [\cdots] =
\frac{1}{\eps} \frac{1}{1+u^2\f{1}{ V_{11} (V^{-1})_{11}}}
	\ ,
\eqx
The determinant corresponding to the second trace in (\ref{main}),
with $(z-{\cal M})$ and $(\bar{z}-{\cal M}^\dagger)$ interchanged 
is obtained by
substituting $V \leftrightarrow V^{-1}$ and hence is given by exactly
the same formula.

Therefore 
\be
\frac{1}{4} \corr{\partial_{\eps} \log \det [\cdots]\; \partial_{\eps}
\log \det [\cdots]} &=& \left< \frac{1}{\eps^2}
 \frac{1}{(1+u^2\f{1}{V_{11} (V^{-1})_{11}} )^2}
\right>  \nonumber \\
&\stackrel{\eps \rightarrow 0}{=}& \left< \pi \delta^{(2)}(z-\lambda_1) 
V_{11} (V^{-1})_{11}\right> \,.
\ee
where we used  the representation for complex Dirac delta 
$\pi \delta^{(2)}(z)= \lim_{\eps \rightarrow 0} \eps^2/(\eps^2 +|z|^2)^2$.

In this way we obtain the important formula
\be
\frac{1}{4}\corr{\partial_{\eps} \log \det [\cdots]\; \partial_{\eps}
\log \det [\cdots]}= \pi \corr{\sum_i V_{ii} (V^{-1})_{ii}
\delta^{2}(z-\lambda_i)} 
\, ,
\label{crucial}
\ee
where we have reinstated the sum over all eigenvalues.
It remains to show that $V_{ii} (V^{-1})_{ii}=<R_i|R_i><L_i|L_i>$.
Using any orthonormal basis $\{e_i\}$ and the decomposition
(\ref{comp}), we see that the linear transformation $U$ satisfying
\eq
U^{-1} {\cal M} U= \sum_k \lm_k |e_k><e_k|
\eqx
can be written explicitly as
\be
U&=&\sum_k |R_k><e_k| \ ,\nonumber \\
V^{-1}&=&U^{\dagger}U=\sum_{k,n} |e_k><R_k|R_n><e_n| \,.
\ee
{}From the last equation we infer 
\be
V^{-1}_{ii}\equiv <e_i|V^{-1}|e_i>=<R_i|R_i> \,.
\ee
Similar reasoning leads to $V_{ii}=<L_i|L_i>$.

This yields
\be
\frac{1}{4} \corr{\partial_{\eps} \log \det [\cdots]\; \partial_{\eps}
\log \det [\cdots]} =\pi \left< \sum_i <L_i|L_i><R_i|R_i> 
\delta^{(2)}(z-\lambda_i) \right> \equiv \pi N O(z)
\label{master}
\ee
which completes the proof of (\ref{main}).

We would like to stress that till now 
we have not used the large-$N$ arguments in
deriving
this formula. Therefore (\ref{main})  is more general than the relation
(\ref{hypothesis}), and could be used as a starting point 
for a systematic study of finite-size effects or microscopic limit.

To complete the proof of relation (\ref{hypothesis}), we observe that,
in the large-$N$ limit, we could use the factorization theorem, 
such that  the l.h.s. of (\ref{main}) splits into the product of averages 
$<{\rm Tr}[...]><{\rm Tr}[...]>$. 

{}From 
the  definitions (\ref{18},\ref{19}), 
the off-diagonal Green's functions have the following
expression:
\be
\GG_{q\bar{q}}&=&\left< -i \lim_{ \eps\rightarrow 0}\frac{1}{N}{\rm Tr}
\frac{\eps}{(z-\MM)(\bar{z}-\MM^{\dagger}) +\eps^2}\right> \nonumber \\
\GG_{\bar{q}q}&=&\left< -i \lim_{ \eps\rightarrow 0}\frac{1}{N}{\rm Tr}
\frac{\eps}{(\bar{z}-\MM^{\dagger})(z-\MM) +\eps^2}\right> 
\ee
Hence 
\be
\frac{\pi}{N}O(z)= -\GG_{q\bar{q}} \GG_{\bar{q}q}
\ee
which completes the proof of equation (\ref{hypothesis}). 

%\vskip 1.5cm
%{\bf Appendix B}
%\vskip .2cm
\section{Generalized Blue's function}

 The generalized Blue's function~\cite{US,ZEENEW1}
 is 
a $2 \times 2$ matrix valued function  defined by
\be
\BB[\GG (\ZZ)] =\ZZ=\left( \begin{array}{cc} z & i\eps \\ i\eps & \bar{z}
                    \end{array} \right) \ , 
\ee
where $\GG$ was defined in Section~2 and $\eps$
 will be eventually set to zero. This is equivalent
to the definition in terms of the self-energy matrix
\be
\BB (\GG )=\Sigma+\GG^{-1} \ ,
\label{genblue}
\ee
where $\Sigma$ is a $2 \times 2$ self-energy matrix expressed as a
function of a matrix-valued Green's function. 
 The addition law for generalized Blue's functions reads 
\be
\ZZ  =\BB_1(\GG)+\BB_2(\GG) -\GG^{-1} .
\label{genadd}
\ee
in analogy to the original construction by Zee~\cite{ZEEOLD} 
for hermitian matrices.

%\vskip 0.9cm 
%\noindent {$\bullet$ \bf Ginibre ensemble}\\
\subsection{Ginibre's ensemble}

Ginibre's ensemble~\cite{GINIBRE} 
could be viewed as a sum of hermitian and anti-hermitian
Gaussian ensembles, with the original width suppressed by $\sqrt{2}$
in relation to the original width of the complex gaussian ensemble.
The generalized Blue's function for the hermitian part
is simply~\cite{US}
\be
{\cal B}_R (\AA)= \frac{1}{2} \AA + \AA^{-1} \,.
\ee
The generalized Blue's function for anti-hermitian part
is~\cite{US} 
\be
{\cal B}_{iR} (\AA)= \frac{1}{2} \tilde{\AA} + \AA^{-1} \ ,
\ee
where we used the notation
\be
\tilde{ \AA}=\arr{-1}{0}{0}{1} \AA \arr{1}{0}{0}{-1} \,.
\ee
The factor $1/2$ comes from the normalization of the width, 
and the extra signs reflect on  the anti-hermitian correlations of the matrix
elements.
The addition law reads now  (in the matrix form)
\be
\ZZ=\GG^{-1}+\frac{1}{2}[ \GG +\tilde{\GG}] \,.
\label{addgin}
\ee
The nontrivial (non-holomorphic solution)
reads
\be
\GG=\arr{\bar{z}}{\sqrt{|z|^2-1}}{\sqrt{|z|^2-1}}{z} \,.
\ee
The domain of eigenvalues is determined by the condition $\GG_{q\bar{q}}=0$,
for which the block structure decouples leading to holomorphic and
anti-holomorphic copies. For this ensemble, this is simply a circle
$|z|^2=1$.
Inside the circle, $\GG_{qq}(z,\bar{z})=\bar{z}$ (upper left corner of $\GG$).
The constant density of eigenvalues follows from Gauss law (\ref{20}).
Outside the circle, the second (holomorphic) solution of (\ref{addgin}) 
is valid, 
giving $G(z)=1/z$.
This reproduces the salient features of Ginibre's ensemble. 

A straightforward generalization~\cite{US} 
 using the  measure~\cite{GIRKO,SOMMERS88}
\be
P(\MM)d\MM \sim \exp \left(-\frac{N}{1-\tau^2} {\rm Tr}(\MM \MM^{\dagger} -\tau
{\rm Re} \MM \MM) \right) d\MM
\label{measure1}
\ee  
leads to the results for the elliptic ensemble 
with 
\be 
\GG_{qq}&=& \frac{\bar{z}-\tau z}{1-\tau^2} \ ,\nonumber \\
\GG_{\bar{q} \bar{q} }&=& \bar{\GG}_{qq} \ ,\nonumber \\
\GG_{\bar{q} q} \GG_{q \bar{q}} &=& 
\frac{(1+\tau^2)|z|^2-2\tau {\rm Re} z^2-(1-\tau^2)^2}{(1-\tau^2)^2} \,.
\ee
Note that the  measure (\ref{measure1}) leads to non-vanishing cumulants 
$\la\MM_{ab}{\overline{\MM}}_{ab}\ra=1/N$
and $\la\MM_{ab} {\MM}_{ba}\ra=\tau /N$. In particular, 
$\tau=-1$ corresponds to anti-hermitian matrices, explaining the flips of the
signs
in the tilted variables above. 

%\noindent
%$\bullet$ {\bf Two level deterministic Hamiltonian plus Ginibre ensemble}\\
\subsection{Two level deterministic Hamiltonian plus Ginibre's ensemble}

Since the deterministic hermitian Green's function 
for the two-level Hamiltonian is
\be
G_D(z)=\frac{1}{2}\left(\frac{1}{z-a} +\frac{1}{z+a}\right)
\ee
the corresponding generalized Green's function
for this  Hamiltonian reads 
\be
\GG_D(\ZZ)=\frac{1}{2}\left[(\ZZ-a{\bf 1}_2)^{-1}+(\ZZ+a{\bf
1}_2)^{-1}\right]
\label{detmat}
\ee 
with ${\bf 1}_2$ denoting   the two-dimensional identity matrix.
Substituting in (\ref{detmat}) $\ZZ \rightarrow \BB_D(\GG)$ we obtain
\be
\GG =\frac{1}{2}\left[(\BB_D(\GG)-a{\bf 1}_2)^{-1}+(\BB_D(\GG)+a{\bf
1}_2)^{-1}\right]
\label{firstblue}
\ee
{}The addition law for generalized  Blue's functions reads 
\be
\BB (\AA)=\BB_D(\AA)+\BB_G(\AA)-\AA^{-1}
\label{bbb}
\ee
where the generalized Blue's function for  Ginibre's 
 ensemble was 
constructed above, i.e. $\BB_G(\AA)=\AA^{-1}+1/2(\AA+\tilde{\AA})$.
Substituting in (\ref{bbb}) $\AA \rightarrow \GG(\ZZ)$ we infer the relation
\be
\ZZ=\BB_D(\GG)+\frac{1}{2}(\GG +\tilde{\GG}) \,\,.
\label{secondblue}
\ee
{}From (\ref{firstblue}) and (\ref{secondblue}) we 
get the final equation
\be
\GG=   \frac{1}{2}\left[ 
(\ZZ-\frac{1}{2}(\GG+\tilde{\GG})-a {\bf 1}_2)^{-1}
+(\ZZ-\frac{1}{2}(\GG+\tilde{\GG})+a {\bf 1}_2)^{-1}\right] \,\, .
\ee
Solving this matrix equation we arrive at
\be
\GG_{\bar{q}q} \GG_{q \bar{q}}&=& |z|^2 +a^2 -\frac{1}{2}(1+\sqrt{1+
4a^2(z+\bar{z})^2}) \ ,\nonumber \\
\GG_{qq}&=& \bar{z}-\frac{2a^2 (z+\bar{z})}{1+\sqrt{1+
4a^2(z+\bar{z})^2}} \ ,\nonumber \\
\GG_{\bar{q}\bar{q}}&=&\bar{\GG}_{qq}  \,.
\ee

%\noindent
%$\bullet${\bf Open chaotic scattering}\\
\subsection{Open chaotic scattering}

Since the addition law for open chaotic scattering was formulated 
by us in previous  publications~\cite{US}, 
we will be brief and refer to the original papers for  details.
The addition law reads
\be
\ZZ=m(1-\Gamma \GG)^{-1}\Gamma +\GG + \GG^{-1}
\label{hakblue}
\ee
where $m=M/N$ and $\Gamma= {\rm diag} \,(-ig,ig)$.
Solution of the matrix equation (\ref{hakblue}) leads to  
\be
\GG_{qq}&=&\frac{x}{2}+\frac{i}{2}\left[\frac{1}{g} +
\frac{m}{y}+\frac{g}{1-gy}
\right] \ , \nonumber\\
\GG_{q \bar{q}}\GG_{q\bar{q} } &=& 
\f{x^2}4 + \f14
	\left( \f{g}{1-gy}+\f{m}y+\f1g \right)^2 - \f1{1-gy} \,.
\nonumber \\
\GG_{\bar{qq}}&=&\bar{\GG}_{qq}
%B
\ee
where  $z=x+iy$.
Gauss law leads to the spectral density~\cite{HAAK}
\be
4\pi \nu(x,y)= {\rm div} \, \vec{E}=
1+\frac{m}{y^2}-\frac{g^2}{(1-gy)^2}
\ee
where $E_x= 2 {\rm Re}\,\, \GG_{qq}$ and  $E_y= - 2 {\rm Im}\,\, \GG_{qq}$. 
For completeness we mention 
that  condition $\GG_{q\bar{q}}=0$ reproduces 
the results 
by Haake, Sommers and coworkers~\cite{HAAK}
 for the boundary  of  eigenvalues
in  open chaotic scattering.

\setlength{\baselineskip}{15pt}


\vspace*{-5mm}
\begin{thebibliography}{50}
\bibitem{SOMMERS}
Y.V. Fyodorov and H.-J.Sommers, 
	J. Math. Phys. {\bf 38} (1997) 1918;
Y.V. Fyodorov, B.A. Khoruzhenko  and H.-J.Sommers,
	Phys. Lett. {\bf A226} (1997) 46. 

\bibitem{EFETOV}
K.B. Efetov,
	Phys. Rev. Lett.{\bf 79} (1997)491;
	Phys. Rev. {\bf B56} (1997) 9630.

\bibitem{HATANONELSON}
N. Hatano and D.R. Nelson,
	Phys. Rev. Lett. {\bf 77} (1997) 370.

\bibitem{US}
R.A. Janik, M.A. Nowak, G. Papp and I. Zahed,
	Nucl. Phys. {\bf B501} (1997) 603;
R.A. Janik, M.A. Nowak, G. Papp, J. Wambach  and I. Zahed,
	Phys. Rev. E. {\bf 55} (1997) 4100.

\bibitem{ZEENEW1}
J. Feinberg and A. Zee, Nucl. Phys. {\bf B501} (1997) 643;
	Nucl. Phys. {\bf B504} (1997) 579.

\bibitem{CHALKER}
J.T. Chalker and B. Mehlig,
	cond-mat/9809090.

\bibitem{ROTTER}
M. M\"{u}ller, F.-M. Dittes, W. Iskra and I. Rotter,
	Phys. Rev. {\bf E52} (1995) 5961.

\bibitem{GINIBRE}
J. Ginibre,
	J. Math. Phys. {\bf 6} (1965) 440.

\bibitem{GIRKO}
V.L. Girko,
	{\it Spectral theory of random matrices} (in Russian), Nauka,
	Moscow (1988) and references therein.

\bibitem{SOMMERS88}
H.-J. Sommers, A. Crisanti, H. Sompolinsky and Y. Stein,
	Phys. Rev. Lett. {\bf 60} (1988) 1895.

\bibitem{WEIDMAUX}
C. Mahaux and H.A. Weidenm\"{u}ller,
	{\it Shell model approach to nuclear reactions}, North Holland, 
	Amsterdam (1969).  

\bibitem{HAAK}
F. Haake et al.,
	Zeit. Phys. {\bf B88} (1992) 359;
N. Lehmann, D. Saher, V.V. Sokolov and H.-J. Sommers,
	Nucl. Phys. {\bf A582} (1995) 223.

\bibitem{ZEL}
V.V. Sokolov and V.G. Zelevinsky,
	Phys. Lett. {\bf B202} (1988) 101;
	Nucl. Phys. {\bf A504} (1989) 562.

\bibitem{NATO}
R.A. Janik, M.A. Nowak, G. Papp and I. Zahed,
	in {\it New developments in quantum field theory}, eds. P. Damgaard
	and J. Jurkiewicz, NATO ASI Series {\bf B}: Physics Vol. 366, p.297. 
	Plenum Press, New York (1998).

\bibitem{BLUESHADES}
R.A. Janik, M.A. Nowak, G. Papp and I. Zahed, 
	Acta Physica Pol. {\bf B28} (1997) 2949.

\bibitem{CHALKERWANG}
J.T. Chalker and Z. Jane Wang,
	Phys. Rev. Lett. {\bf 79} (1997) 1797.
 
\bibitem{ZEEOLD}
A. Zee,
	Nucl. Phys. {\bf B474} (1996) 726.
\end{thebibliography}
\end{document}